# Attention Hybrid Variational Net for Accelerated MRI Reconstruction

Guoyao Shen [1], Boran Hao [2], Mengyu Li [1], Chad W. Farris [3], Ioannis Ch. Paschalidis [2], Stephan W. Anderson [3], and Xin Zhang*[1]

[1] Department of Mechanical Engineering and the Photonics Center, Boston University, Boston, MA, 02215 USA

[2] Department of Electrical and Computer Engineering, Boston University, Boston, MA, 02215 USA

[3] Boston Medical Center and Boston University Chobanian & Avedisian School of Medicine, Boston, MA, 02118 USA

**\* Corresponding author:** Prof. Xin Zhang; email: xinz@bu.edu

**Acknowledgments:** This work was supported by the Rajen Kilachand Fund for Integrated Life Science and Engineering. We would like to thank the Boston University Photonics Center for technical support. GS acknowledges the BUnano Cross-Disciplinary Fellowship from Boston University.

**Abstract**

The application of compressed sensing (CS)-enabled data reconstruction for accelerating magnetic resonance imaging (MRI) remains a challenging problem. This is due to the fact that the information lost in k-space from the acceleration mask makes it difficult to reconstruct an image similar to the quality of a fully sampled image. Multiple deep learning-based structures have been proposed for MRI reconstruction using CS, both in the k-space and image domains as well as using unrolled optimization methods. However, the drawback of these structures is that they are not fully utilizing the information from both domains (k-space and image). Herein, we propose a deep learning-based attention hybrid variational network that performs learning in both the k-space and image domain. We evaluate our method on a well-known open-source MRI dataset and a clinical MRI dataset of patients diagnosed with strokes from our institution to demonstrate the performance of our network. In addition to quantitative evaluation, we undertook a blinded comparison of image quality across networks performed by a subspecialty trained radiologist. Overall, we demonstrate that our network achieves a superior performance among others under multiple reconstruction tasks.

**Introduction**

Magnetic resonance imaging (MRI) is a powerful tool in clinical medicine and research settings, from diagnosing knee injuries to studying brain disease. However, one major challenge for MRI is the long data acquisition time, ultimately limiting patient access to this diagnostic tool. Furthermore, the long image acquisition times also lead to patient discomfort, worsen potential claustrophobia, increased chance of patient motion and degraded images that reduce diagnostic utility. In order to reduce the scan time and improve the efficiency, compressed sensing (CS) techniques [1, 2] that enable less measurement data in k-space in order to reconstruct the final image have been developed and successfully applied to MRI. However, these under-sampling strategies come at a cost in image quality with resultant image blurring and aliasing artifacts which can significantly influence the diagnostic yield of MRI.

Recently, thanks to achievements in deep learning, especially the successful application of deep convolutional neural networks (DCNN) in multiple imaging tasks [3], a new paradigm has been realized in which image reconstruction can be accomplished by exploiting the latent representation within the neural network [4, 5, 6, 7]. Importantly, these deep learning-based strategies also provide an additional benefit whereby their network structure requires no or minimal modification before they may be applied to different tasks. For a typical CS reconstruction task in MRI, there are two major domains of relevance: the k-space domain and the image domain. Data under-sampling typically occurs in k-space, such that a mask is applied, and k-space is incompletely sampled. If Fourier transformation is then applied to this under-sampled k-space data with all the masked portions zero-filled, this is referred to as zero-filled MRI (ZF-MRI). The overall quality of ZF-MRI relies on the degree of under-sampling and the pattern of the k-space mask [1, 8]. There are ongoing efforts focused on reconstructing a high-quality image using as input the under-sampled k-space or the ZF-MRI.

To this end, multiple deep learning structures have been proposed to leverage the CS reconstruction task and improve the ultimate image quality. U-Net is a well-known structure which was proposed to handle image segmentation tasks [9], yet later was reported also to be feasible for MRI reconstruction [10, 11]. The U-Net structure focused on the reconstruction in the image domain, employing a ZF-MRI as input and reconstructing an unaliased image. It has also been shown that employing a data consistency (DC) layer [12] serves to improve the performance of deep convolutional neural network (DCNN) in the image domain due to the reuse of the visible portions of k-space. Furthermore, with an increase in interest in the attention mechanism in DCNNs, spatial- and channel-wise attention approaches [13, 14] have also been reported to provide additional benefits in reconstruction performance.

In contrast to methods focused on the image domain, automated transform by manifold approximation (Automap) [15] is a method focused on learning the domain transformation in order to perform reconstruction directly from the k-space domain. The fully connected layers encode the transformation of the complex data into image space, followed by a convolution-deconvolution structure that provides the reconstruction of the final image. Even though this method provides a prototype of learning from the combination of both the k-space and image space domains, one major disadvantage of this method is its sensitivity to image size. In the Automap technique, the dimensions of the fully connected layers are closely related to the image shape, making it difficult to train for a larger image.

Unrolled optimization-based structures, on the other hand, are a type of network that divides the overall reconstruction task into multiple steps, where each step contains one or multiple sub-networks [16]. Variational net (VarNet) [17, 18] is one such unrolled optimization-based structure that utilizes not only the under-sampled k-space data, but also the mask and sensitivity maps during intermediate states.

W-Net (Double U-Net) is a hybrid network which performs reconstruction in both k-space and image domains [19]. The network consists of two U-Net structures and an inverse Fourier transformation layer connecting them. It has been shown that this dual-domain learning structure provides a better reconstruction performance compared to a single-domain learning structure. Yet the W-Net lacks in the k-space learning efficiency as the k-space domain U-Net holds a small weight in the combined loss function, putting more weight in the image-domain learning.

In this work, we seek to leverage and further develop a CS reconstruction method. More specifically, we build an attention hybrid variational network (AttHybrid-VarNet) that benefits from the superior k-space reconstruction ability and an image-domain refinement network to further improve the image quality. Furthermore, spatial- and channel-wise attention also enables the convolutional module to further fine tune the weights for different channels and regions in the feature maps according to the attention scores. We compare our architecture with multiple CS reconstruction networks over open-source and clinical imaging datasets. We also provide a blinded radiologist evaluation of image quality of our methods compared to other typical reconstruction networks. Ultimately, we demonstrate that our network achieves a superior performance among others under multiple reconstruction setups.

**Materials and Methods**

*1. End-to-end variational network*

Variational networks have shown superior performance for MRI reconstruction tasks. Consider a MR image acquiring measurement:

$$k = \mathcal{F}(x) + e \tag{1}$$

where $x$ is the underlying image and $k$ is the k-space measurement, $\mathcal{F}$ is the Fourier transform operator, $e$ stands for the measurement noise. In an accelerated MRI acquisition case, $\widetilde{k} = Mk$, $M$ is the binary under-sampling mask applied to the k-space and $\widetilde{k}$ denotes the under-sampled k-space measurement.

An estimation of the underlying image can be solved using the following optimization:

$$\hat{x} = argmin_x \frac{1}{2}\left\|A(x) - \widetilde{k}\right\|^2 + \lambda \psi(x) \tag{2}$$

$$x^{t+1} = x^t - \eta^t \left(A^*\bigl(A(x) - \widetilde{k}\bigr) + \lambda \phi(x^t)\right) \tag{3}$$

Here $A$ is a linear operator that applies sensitivity maps, perform 2D Fourier transform and then under-samples the k-space. $A^*$ is the Hermitian of $A$. $\psi$ is a regularization term, $\phi$ is the gradient of $\psi$ with respect to $x$. $\eta^t$ is the learning rate.

Variational network uses a small convolutional neural network (CNN) for each gradient step in equation 3:

$$x^{t+1} = x^t - \eta^t \left(A^*\bigl(A(x) - \widetilde{k}\bigr) + CNN(x^t)\right) \tag{4}$$

End-to-end variational network (E2E-VarNet) [18] shows that the aforementioned gradient update can be further formulated as:

$$k^{t+1} = k^t - \eta^t M(k^t - \widetilde{k}) + G(k^t) \qquad (5)$$

where $G(k^t) = \mathcal{F} \circ \mathcal{E} \circ CNN(\mathcal{R} \circ \mathcal{F}^{-1}(k^t))$. $\mathcal{E}$ and $\mathcal{R}$ are expand and reduce operator. The expand operator take image and sensitivity maps $S_i$ and output corresponding images, while the reduce operator combine individual coil images: $\mathcal{E}(x) = (x_1, \dots, x_N) = (S_1 x, \dots, S_N x)$, $\mathcal{R}(x_1, \dots, x_N) = \sum_i S_i^* x_i$. For variational network, the cascading CNN in each step can be a small U-Net structure [9].

## 2. Dual-domain learning

Compared to one network performing reconstruction only in the k-space domain or image domain, dual-domain learning has shown overall better image qualities [19]. A dual-domain learning structure includes a k-space reconstruction network and an image domain reconstruction network linked by an inverse Fourier transform layer. In our setup, we take the benefit of the superior performance from variational network and have a simple U-Net for further image domain refinement. Our network has a balanced weight in both domains so that each sub-network gets properly trained.

## 3. Spatial- and channel-wise attention

For U-Nets in both domains, we adopted the spatial and channel wise attention mechanism [14]. Suppose the feature map for a certain intermediate CNN block has the shape: $X \in \mathbb{R}^{H \times W \times C}$, where $H \times W$ is the shape of the map and $C$ is the number of channels of the feature map. For k-space input, we treated the real part and the imaginary part as two channels stacked, where each channel has real values inside.

The feature map can be seen as a combination of all channels $X = [x_1, x_2, \ldots, x_C], x_i \in \mathbb{R}^{H \times W}$. Channel-wise attention is achieved by first squeezing the spatial dimensions: taking a global average for each feature map:

$$z_k = \frac{1}{H \times W} \sum_i^H \sum_j^W x_k(i,j)$$

Vector $z_k \in \mathbb{R}^{1 \times 1 \times C}$ contains the global spatial information of the feature map. It is then passed to a fully connected network followed by a sigmoid activation function to learn the channel-wise attention:

$$s_c = \sigma\left(W_{fc}(z)\right) \tag{6}$$

Here we used the ReLU function as the intermediate activation for the fully connected network $W_{fc} \in \mathbb{R}^{C \times \frac{C}{2} \times C}$. $X$ was element-wisely multiplied by $s_c$ as the output:

$$y_c = s_c \odot X$$

Spatial-wise attention tries to learn a score for each pixel considering all channels. Indexing the same feature map spatially, we have $X = [x_{1,1}, x_{1,2}, \ldots, x_{i,j}, \ldots, x_{H,W}], x_{i,j} \in \mathbb{R}^{1 \times 1 \times C}$. The spatial-wise attention is learned by applying a pixel-wise convolution $W_{conv} \in \mathbb{R}^{1 \times 1 \times C \times 1}$, followed by a sigmoid function:

$$s_s = \sigma(W_{conv} \circledast X) \tag{7}$$

where $s_s \in \mathbb{R}^{H \times W \times 1}$. And the output is the element-wise multiplication of $s_s$ and $X$:

$$y_s = s_s \odot X$$

As shown above, the channel-wise attention squeezes the spatial dimension and learns an importance factor for each channel, while the spatial-wise attention squeezes the channel dimension and learns an importance factor for each pixel. The final output is the element-wise max-out of these two attentions:

$$y = max(y_c, y_s) \tag{8}$$

### 4. Attention hybrid variational network

Our Attention hybrid variational network (AttHybrid-VarNet) structure is shown in Figure 1. It consists of an end-to-end VarNet (E2E-VarNet) for k-space domain learning and a refinement network for image domain learning. Similar to the E2E-VarNet, our Atthybrid-VarNet uses k-space quantities rather than image-space quantities only and requires no pre-training and fine-tuning or parameter freezing process, making it an end-to-end model. We use a weighted combination of normalized root mean squared errors (NRMSE) for both the k-space domain and image domain learning:

$$L = NRMSE(x, \hat{x}_{intermediate}) + \alpha NRMSE(x, \hat{x})$$

$$NRMSE(x, \hat{x}) = \frac{\sqrt{MSE(x, \hat{x})}}{max(x) - min(x)}$$

where $\hat{x}_{intermediate}$ and $\hat{x}$ are the intermediate and final reconstruction of the fully sampled reconstruction image $x$. $\alpha$ is a weighting factor for the image-domain refinement. In our setup, we simply set $\alpha = 1$ for a balanced dual-domain learning.

### 5. Network, training and dataset details

The image domain refinement network design can be rather flexible. In our setup, we used a U-Net structure for simplicity. We followed the structure proposed in [9] with 32 channels for the shallowest layer's convolutional block. In addition, we applied an attention layer at the end of each depth's convolution block for the U-Net in the image domain refinement network and the k-space domain VarNet. Figure 2 illustrates the structure of the attention layer. We compared our AttHybrid-VarNet against U-Net, W-Net [19] and E2E-VarNet. We used Pytorch [20] to implement our network. During the training process we used the Adam optimizer [21] with a learning rate of 0.001 for all the models mentioned.

We evaluated our model on a large-scale open-source MRI dataset fastMRI and a dataset derived from patients imaged at our tertiary care hospital and diagnosed with stroke. For the fastMRI dataset, each slice was center cropped to $320 \times 320$ for evaluation. We ran tests on both the knee and brain 4-fold and 8-fold reconstruction tasks with center fraction for 0.08 in 4-fold and 0.04 in 8-fold [11]. Due to its large data scale, we randomly selected 521 and 131 cases from the brain dataset for training and testing. For the knee dataset, we used the full single-coil training and validation set for training and testing.

In addition to the fastMRI dataset, we employed a retrospective clinical dataset of patients diagnoses with stroke at our institution. In this case, our data was collected from both 1.5T and 3T MRI scanners (Philips Healthcare) including 243 patients. In all cases, the b1000 images from a diffusion weighted sequence (DWI) were employed for analysis and acquired using a 2D acquisition with a slice thickness of 5mm. The acquisition parameters for the 1.5T scans are as follows: $TE/TR = 68.8ms/4183ms, FOV = 240 \times 240mm, pixel\ size = 0.94 \times 0.94mm$. For 3T scans: $TE/TR = 86.0ms/4105ms, FOV = 230 \times 230mm, pixel\ size = 1.2 \times 1.2mm$. Our b1000 dataset included 7389 slices in total, with 1650 slices containing stroke. We randomly split it using the ratio of 80%/20%, leading to 5904 slices for training and 1485 slices for testing. We resized all slices to $256 \times 256$ and performed more challenging 20-fold and 30-fold reconstruction tasks with two-dimensional Gaussian sampling. Our training and test set consisted of 5904 and 1485 slices, respectively. We conducted a blinded image quality test using a single board-certificated radiologist with subspecialty certification in neuroradiology (C.W.F.) for evaluation of the quality of the reconstructions of four models. We randomly extracted 100 slices of 79 patients from the test set. For each slice sample, the radiologist was given four candidate reconstructions and asked to rank their preference from 1 (most preferable) to 4 (least preferable).

To evaluate the image quality of our reconstructions, we report peak signal-to-noise ratio (PSNR) and structural similarity (SSIM) with respect to fully sampled ones. Furthermore, we performed a blinded test on our b1000 dataset for image quality evaluation. The test sample slices were extracted randomly from the test set with adequate brain area and had the neuroradiologst label preference priorities from 1 to 4 with 1 representing the most preferable and 4 for the least preferable. The priority score is then given by:

$$S_{priority} = \frac{N_{candidates} + 1 - \frac{1}{N}\sum_{i=1}^{N} p_i}{N_{candidates}} \tag{9}$$

where $N_{candidates} = 4$ is the number of candidates for each slice being evaluated. $N$ is the number of samples (slices). $p_i$ stands for the priority rank. Higher priority score $S_{priority}$ means its more preferred in the blind test.

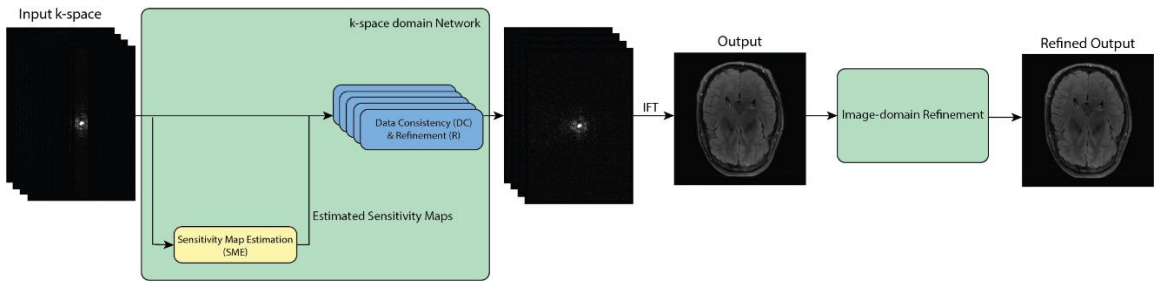

**Figure 1.** The overall structure of our attention hybrid variational network (AttHybrid-VarNet). It consists of a variational network for k-space domain learning and an image domain refinement network. The k-space domain network includes a sensitivity map estimation module and multiple data consistency and refinement blocks for unrolled optimization learning. The image domain refinement network takes the intermediate reconstruction from the k-space domain network and give the refined output.

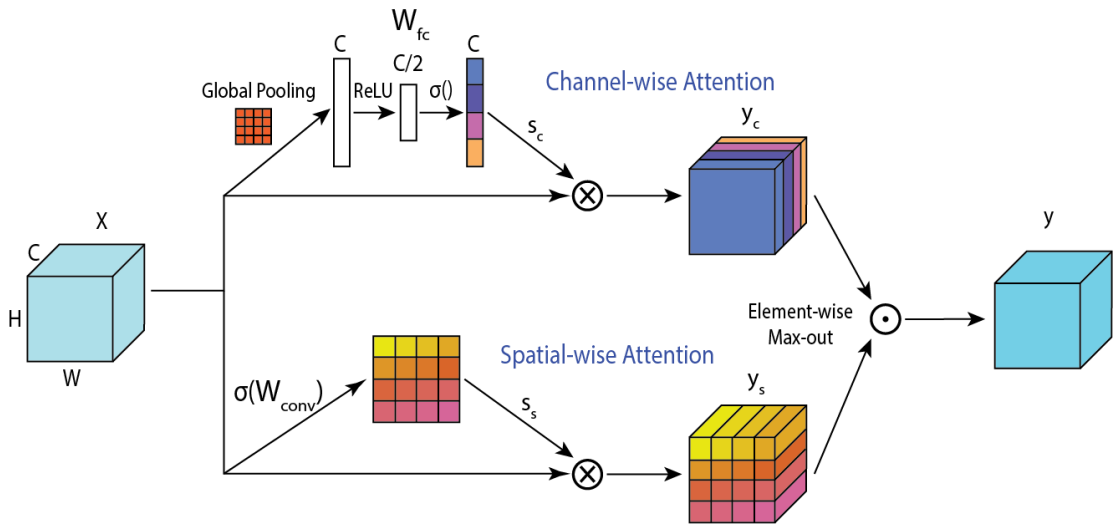

**Figure 2.** Detailed structure of the attention layer. It includes two attention paths: channel-wise attention and spatial-wise attention. The outputs $y_c$ and $y_s$ from two paths are followed by an element-wise max-out operation to get the final output $y$.

# Results

## 1. Performance on fastMRI dataset

The fastMRI brain dataset contains 8336 slices for training and 2096 slices for testing. Table 1 demonstrates the evaluation metrics. For the 4-fold acceleration task in brain reconstruction, our model achieved an overall $PSNR = 40.92, SSIM = 0.9577$. In the 8-fold brain reconstruction task, our model achieved an overall $PSNR = 37.03$, $SSIM = 0.9365$. As for the knee reconstruction tasks, for 4-fold acceleration, our model had an overall $PSNR = 31.09, SSIM = 0.6901$. And $PSNR = 29.49, SSIM = 0.6197$ for the 8-fold acceleration test. Reconstruction samples for the fastMRI dataset are shown in Figure 3. The size of the image is $320 \times 320$ pixels. Red boxes depict windows with the size of $90 \times 90$ pixels with its enlarged details showing on the upper-right side and its corresponding error maps on the lower-right side. All the error maps are unit normalized and enhanced three times for better demonstration. All models performed well in the 4-fold test as the acceleration factor is small. For the 8-fold test, our model yielded cleaner error maps for both the brain and knee reconstruction. Interestingly, although E2E-VarNet achieved better evaluation metrics than W-Net in the 8-fold tasks, their enlarged area depicted similar level of error.

## 2. Performance on b1000 dataset

Table 2 illustrates the evaluation metrics of the b1000 dataset. Our dataset included 5904 slices for training and 1485 slices for testing. To demonstrate the effectiveness of our model, we ran 20- and 30-fold acceleration reconstruction for the b1000 dataset. For the 20-fold acceleration, our model achieved an overall $PSNR = 36.32$, $SSIM = 0.9199$. For 30-fold tasks, our model achieved an overall $PSNR = 33.70, SSIM = 0.8882$.

Table 3 shows the blinded image quality test results for image quality preference over 100 sample slices from the b1000 dataset. The raw priority rank was calculated by summing up 100 priority ranks for each candidate model as determined by the radiologist. Then, the priority score was calculated according to equation 9. Higher priority scores represent the more preferred reconstruction from the corresponding candidate model choices. Our model achieved an overall score of 0.89 and 0.90 for the 20- and 30-fold tests, respectively.

Figure 4 depicts sample reconstructions for the b1000 dataset. The size of the image is $256 \times 256$ pixels with red boxes highlights a window of size $70 \times 70$ pixels. Enlarged image details and its corresponding error maps are shown on the right side of the image. All the error maps are unit normalized and enhanced three times for better demonstration. The improvement from U-Net to E2E-VarNet is even more noticeable under higher acceleration factors thanks to the unrolled optimization structure. The comparison between the W-Net and U-Net illustrates the benefits of performing dual-domain learning. Our model benefits from both and gives cleaner error maps in samples with or without stroke.

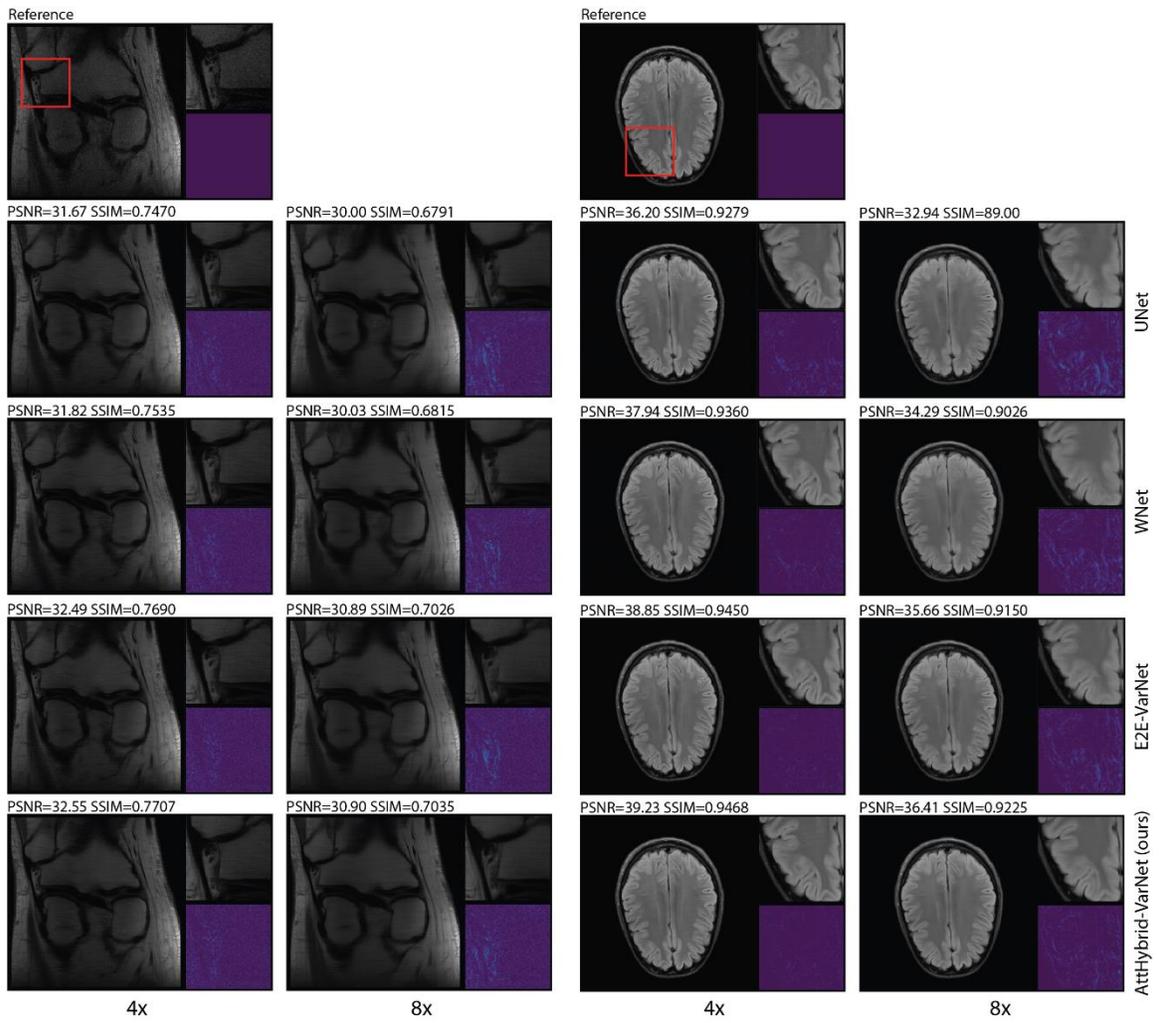

**Figure 3.** Reconstruction samples from the fastMRI dataset. Red boxes in the fully sampled reference scans highlight areas for the enlarged patches and the corresponding error maps. Evaluation metrics for each model and acceleration factor are labeled together with the images.

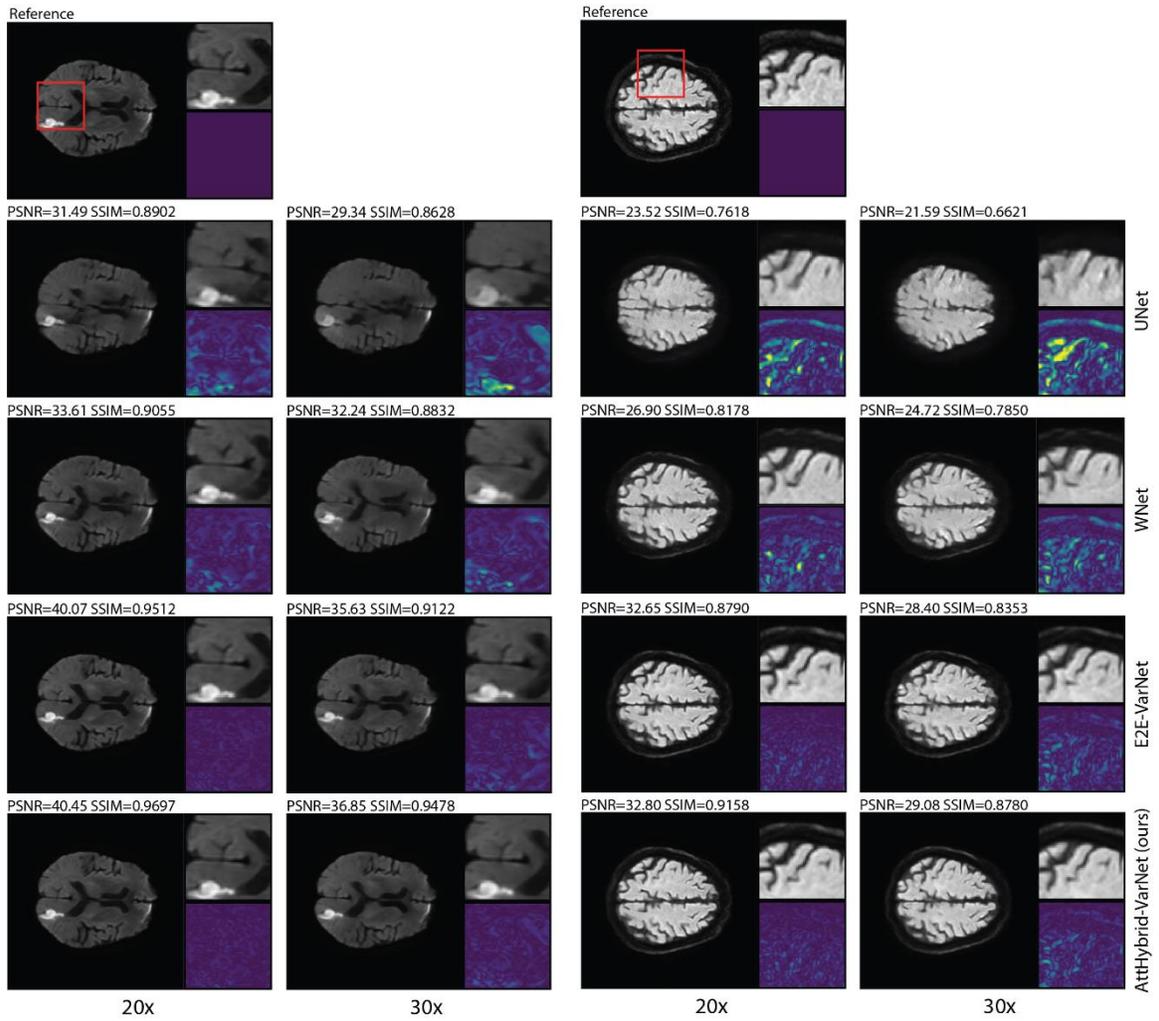

**Figure 4.** Reconstruction samples from the b1000 dataset. Red boxes in the fully sampled reference scans highlight areas for the enlarged patches and the corresponding error maps. Evaluation metrics for each model and acceleration factor are labeled together with the images.

**Table 1.** Summary of results on the fastMRI dataset for different models over multiple acceleration factors and reconstruction tasks.

| | Brain | | | |
|---|---|---|---|---|
| | 4x | | 8x | |
| | PSNR | SSIM | PSNR | SSIM |
| U-Net | 38.08 | 0.9458 | 34.10 | 0.9176 |
| W-Net | 39.52 | 0.9505 | 34.87 | 0.9200 |
| E2E-VarNet | 40.81 | 0.9568 | 36.75 | 0.9340 |
| AttHybrid-VarNet | **40.92** | **0.9577** | **37.03** | **0.9365** |
| | Knee | | | |
| | 4x | | 8x | |
| U-Net | 30.45 | 0.6777 | 28.55 | 0.6038 |
| W-Net | 30.61 | 0.6808 | 28.73 | 0.6060 |
| E2E-Varnet | 31.07 | 0.6899 | 29.48 | 0.6187 |
| Atthybrid-VarNet | **31.09** | **0.6901** | **29.49** | **0.6197** |

**Table 2.** Summary of results on the b1000 dataset for different models over multiple acceleration factors and reconstruction tasks.

| | 20x | |
|---|---|---|
| | PSNR | SSIM |
| U-Net | 29.67 | 0.8230 |
| W-Net | 31.73 | 0.8434 |
| E2E-VarNet | 36.25 | 0.9039 |
| AttHybrid-VarNet | **36.32** | **0.9199** |
| | 30x | |
| U-Net | 28.45 | 0.7920 |
| W-Net | 30.50 | 0.8193 |
| E2E-VarNet | 33.37 | 0.8642 |
| AttHybrid-VarNet | **33.70** | **0.8882** |

**Table 3.** Raw summed priority rank and priority score for b1000 reconstructions over multiple models evaluated by radiologists.

| | Raw Priority Rank | |
|---|---|---|
| | 20x | 30x |
| U-Net | 396 | 396 |
| W-Net | 296 | 301 |
| E2E-VarNet | 164 | 161 |
| AttHybrid-VarNet | **144** | **142** |
| | Priority Score | |
| U-Net | 0.26 | 0.26 |
| W-Net | 0.51 | 0.50 |
| E2E-VarNet | 0.84 | 0.85 |
| AttHybrid-VarNet | **0.89** | **0.90** |

**Discussion**

Prior works on accelerated MRI reconstruction have typically focused on single-domain learning. Previous dual-domain learning structures such as W-Net have a similar structure modality in both domains. In contradistinction, in this study, we focused on building a dual-domain learning structure with different modalities in each domain. Unrolled optimization structures such as the variational networks have been reported to be more suitable for k-space reconstruction [17, 18], which inspired our use of an attention hybrid variational network. Notably, the choice for the image domain refinement network can be flexible and with larger and finer structures, one can anticipate further improvements in performance. Furthermore, one also needs to consider the training process according to the structure design. Larger refinement networks in the image domain can lead to a longer and more challenging training process. This can be mitigated, however, by using fine-tuning with an out-of-box pre-trained model checkpoint. Nevertheless, we kept the structure in the image-domain as simple as possible such that our model can be trained in an end-to-end manner.

Apart from the widely used fastMRI dataset, we tested our model on our stroke dataset and performed an evaluation by a single subspecialty trained radiologist. Rather than using the small acceleration factors recommended by the fastMRI dataset, we tested our model with more challenging cases to demonstrate its effectiveness. When compared to the results from the fastMRI dataset, the difference among models is even more noticeable. In fact, the priority score of the E2E-VarNet when compared to U-Net and W-Net further validate the effectiveness of a variational net for k-space learning. Ultimately, our model demonstrated an overall superior performance in numerical metrics and blinded imaged analysis.

Several limitations of our study are of note. First, as an image regression model, smoothness in the reconstructed images was an expected effect, which is inherited in the loss function and training procedure. In fact, loss functions such as mean squared error, absolute value error or structural similarity error all lead to smoothness. This can be mitigated by dithering the image by a small amount of Gaussian noise in order to preserve the sharpness. More delicate loss functions and training processes would be an interesting topic for future consideration. Second, similar to previous models, our model is task specific. One large topic for future consideration would be designing and training a universal reconstruction network. Recent developments in generative AI have shown promising results in the common image contents field. One challenge in developing a universal reconstruction model would be the requirement for an even larger dataset. Another related question would relate to the fashion by which to properly encode the reconstruction task for a universal model. Nevertheless, this avenue of inquiry would be an interesting topic for future development efforts.

In summary, we report the development of an attention hybrid variational network for accelerated MRI reconstruction. Our model benefits from an unrolled optimization structure and dual-domain learning. We tested our model on a large-scale dataset and then validated our model on a clinically-relevant stroke database from our own institution. We performed numerical evaluation and blinded image quality analyses to demonstrate the effectiveness of our model. In future studies, we hope this work can serve as a reference for cross-domain multi-modality network for image reconstruction.